\documentclass{mn2e} 
\usepackage{psfig}

\title[Cluster Temperature Maps]{Simulated X-ray Cluster Temperature Maps}
\author[L.I. Onuora, S.T. Kay and P.A. Thomas]
{Lesley I. Onuora,\thanks{E-mail: lonuora@pact.cpes.susx.ac.uk} 
Scott T. Kay and Peter A. Thomas\\
Astronomy Centre, University of Sussex, Falmer, Brighton BN1 9QJ}

\date{\today}

\begin{document}

\maketitle

\begin{abstract}
Temperature maps are presented of the 9 largest clusters in the mock
catalogues of Muanwong et al. for both the {\it Preheating} and {\it
Radiative} models.  The maps show that clusters are not smooth,
featureless systems, but contain a variety of substructure which
should be observable. The surface brightness contours are generally
elliptical and features that are seen include cold clumps, hot spiral
features, and cold fronts. Profiles of emission-weighted temperature,
surface brightness and emission-weighted pressure across the surface
brightness discontinuities seen in one of the bimodal clusters are
consistent with the cold front in Abell 2142 observed by Markevitch et
al.

\end{abstract}


\section{INTRODUCTION}
Observations of X-ray clusters of galaxies over the last few years
have shown that clusters are not the smooth, featureless systems they
were expected to be.  X-ray surface brightness and temperature
observations indicate the presence of substructure and support the
view that cluster formation occurs through the infall and merger of
subclusters.

Merger shocks would be expected to occur in such a scenario
(e.g. Markevitch \& Vikhlinin, 2001; Markevitch et al., 2002) but
additionally {\it Chandra} observations have revealed a new phenomenon
of `cold fronts' (Markevitch et al, 2000; Forman et al., 2001;
Mazzotta, Fusco-Femiano \& Vikhlinin, 2002). At a cold front, the
entropy jump across the sharp gas density discontinuity is in the
opposite sense to that expected for a shock, with the high surface
brightness side of the dense edge corresponding to low temperature.

The variety of features that have been observed in the clusters mapped
so far are of great interest since they may shed light on the physical
processes which are occurring as well as containing information on the
stage, geometry, scale and velocity of the mergers.  A number of
authors have carried out simulations of controlled, single mergers
(see Ritchie \& Thomas 2002 for a review). For example, Ritchie \&
Thomas (2002) carried out high-resolution simulations of the merger of
idealized clusters containing both dark matter and gas. They studied
the effect on the observable properties of clusters of single head-on
and off-centre mergers between both equal and unequal mass objects.  A
sequence of maps of emission-weighted temperature with superimposed
X-ray surface brightness and velocity fields for mergers between equal
mass systems showed the compression and shocking of the gas as the
merger progressed. Recently Nagai \& Kravtsov (2002) carried out a
detailed study of cold fronts in high resolution simulations of two
clusters forming in different Cold Dark Matter models (standard CDM
and $\Lambda$CDM). Their results indicate that cold fronts are
probably fairly common but are non-equilibrium transient phenomena.

In this paper, we present preliminary results from temperature maps of
an ensemble of clusters that form within a cosmological simulation,
already shown to reproduce the observed X-ray scaling relations at low
redshift (Muanwong et al.~2002). This way, we are able to directly
assess the range of substructure present in the cluster population.
In common with observations, we see significant temperature
fluctuations in the hot gas, even when there is little information
present in the surface brightness distribution. In particular, we
discuss the presence of a cold front in one of our bimodal clusters,
which has properties consistent with the cold front in Abell 2142
(Markevitch et al.~2000).

\section{The simulations}

Simulation data were generated using a parallel version of the {\sc
hydra} $N$-body/hydrodynamics code (Couchman, Thomas \& Pearce 1995;
Pearce \& Couchman 1997) on the Cray T3E supercomputer at the
Edinburgh Parallel Computing Centre. Full details of the simulations
have already been presented elsewhere (Thomas et al. 2002; Muanwong et
al.  2002, hereafter M2002), so we summarize pertinent details only.

Results are presented for a flat, low-density cosmology
($\Omega_{0}=0.35, \Omega_{\Lambda} = 0.65, h=0.71, \sigma_8=0.9$),
using $160^3$ each of gas and dark matter particles within a box of
comoving length, $100h^{-1}$Mpc.  We consider 2 models: the {\it
Radiative} model includes radiative cooling of the gas, and the {\it
Preheating} model for which we additionally preheat the gas by raising
its specific thermal energy by 1.5 keV per particle at $z=4$.  In both
models, we adopted a time-dependent global metallicity,
$Z=0.3(t/t_0)Z_{\odot}$, where $t_0 \sim 13$Gyr is the current age of
the universe.  As discussed in M2002, both models reproduce the X-ray
cluster scaling relations, although the {\it Radiative} model contains
a significantly higher cooled fraction than the {\it Preheating} model
(15 per cent, as opposed to $\sim 0.5$ per cent).  In the former model,
cooling is limited by numerical resolution.  The cooled gas fraction
is larger than that determined by Balogh et~al.~(2001), using results
from the 2MASS and 2dF galaxy surveys (Cole et al. 2001), but in close
agreement with the observed value from the SDSS (Blanton et al. 2001).

Simulated cluster catalogues were produced using the same method as
described in M2002.  In summary, clusters were identified as clumps of
particles within spheres of average overdensities compared to the
comoving critical density, centred on the position of the densest dark
matter particle. For this paper, we adopt an overdensity of 1000; the
associated radius, $R_{1000}$ is approximately half the size of the
virial radius, and is comparable to the extent currently probed by
{\it Chandra}.
     
\section{The temperature and surface brightness maps}

Maps of emission-weighted temperature were made for each
of the 9 most massive clusters ($M_{1000} = 1-4 \times 10^{14} h^{-1}
\mathrm{M_{\odot}}$) in the {\it Preheating} catalogue, and for the
corresponding clusters in the {\it Radiative} catalogue.  Each map was
produced by first locating all hot ($T>10^5$K) gas particles within a
cube of length, $2R_{1000}$ along each side, centred on the cluster of
interest.  The desired quantity was then smoothed onto a 3D array 
using the SPH interpolation method (e.g.~Monaghan 1992)
\begin{equation}
A({\bf r}_j)
= { \sum_i A_i \, w_i \, \mathcal{W}(\Delta r_{ij},h_i)
                 \over \sum_i w_i \, \mathcal{W}(\Delta r_{ij},h_i) },
\label{eqn:smooth}
\end{equation}
where 
\begin{equation}
\mathcal{W}(\Delta r_{ij},h_i)=
{ W(\Delta r_{ij},h_i) \over 
\sum_k \, W(\Delta r_{ik},h_i)}.
\label{eqn:wnorm}
\end{equation}

The sums $i$ extend over all particles and $k$ over all voxels.  $A_i$
is the value of the quantity for particle $i$ at position ${\bf r}_i$,
${\bf r}_j$ is the centroid position of voxel $j$, 
$\Delta r_{ij} = | {\bf r}_j - {\bf r}_i |$, $w_i$ is a weight factor,
$W(\Delta r, h)$ is the same SPH smoothing kernel used by {\sc hydra} 
(Thomas \& Couchman 1992) and $h_i$ is the smoothing length of particle
$i$ ($W \rightarrow 0$ as $\Delta r \rightarrow 2h$).
Equation~\ref{eqn:wnorm} normalizes the contribution of each particle
to be $A_i$ when summed over all voxels. 
For all mass-weighted quantities, $w_i = m_i$ and for emission-weighted 
quantities, $w_i = m_i n_i \Lambda(T_i,Z)$, for particle $i$ with
mass $m_i$, density $n_i$, temperature $T_i$ and X-ray 
emissivity, $\epsilon_i = n_i^2 \Lambda(T_i,Z)$; the cooling 
function, $\Lambda(T,Z)$ is the same function adopted for our 
simulations, using tables published in Sutherland \& Doptia (1993).
The smoothed distribution was then projected onto a $64 \times 64$ 
grid, with the width of each pixel being
approximately 7 to 10 $h^{-1}\mathrm{kpc}$, smaller than the
gravitational softening length used in the simulations (25
$h^{-1}\mathrm{kpc}$).

\begin{figure*}
\centerline{\psfig{figure=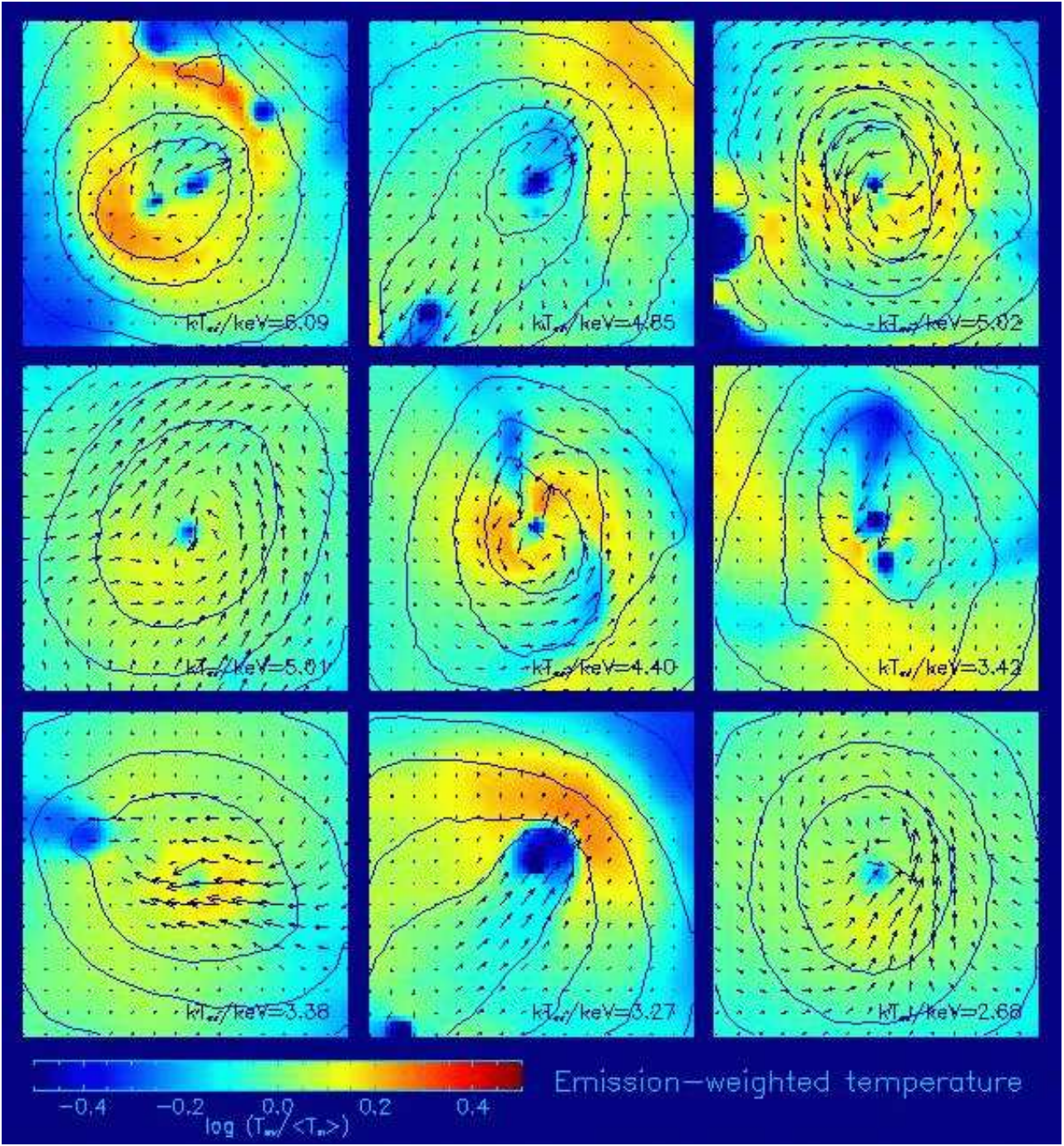,width=17.0cm,angle=0}}
\caption{Temperature maps of the 9 largest clusters in the {\it
Preheating} run with surface brightness contours and velocity vectors
overlaid.  We label the clusters 1--9 in order of virial temperature,
starting from the top-left and reading across then down.} 
\label{fig:phmap}
\end{figure*}

\begin{figure*}
\centerline{\psfig{figure=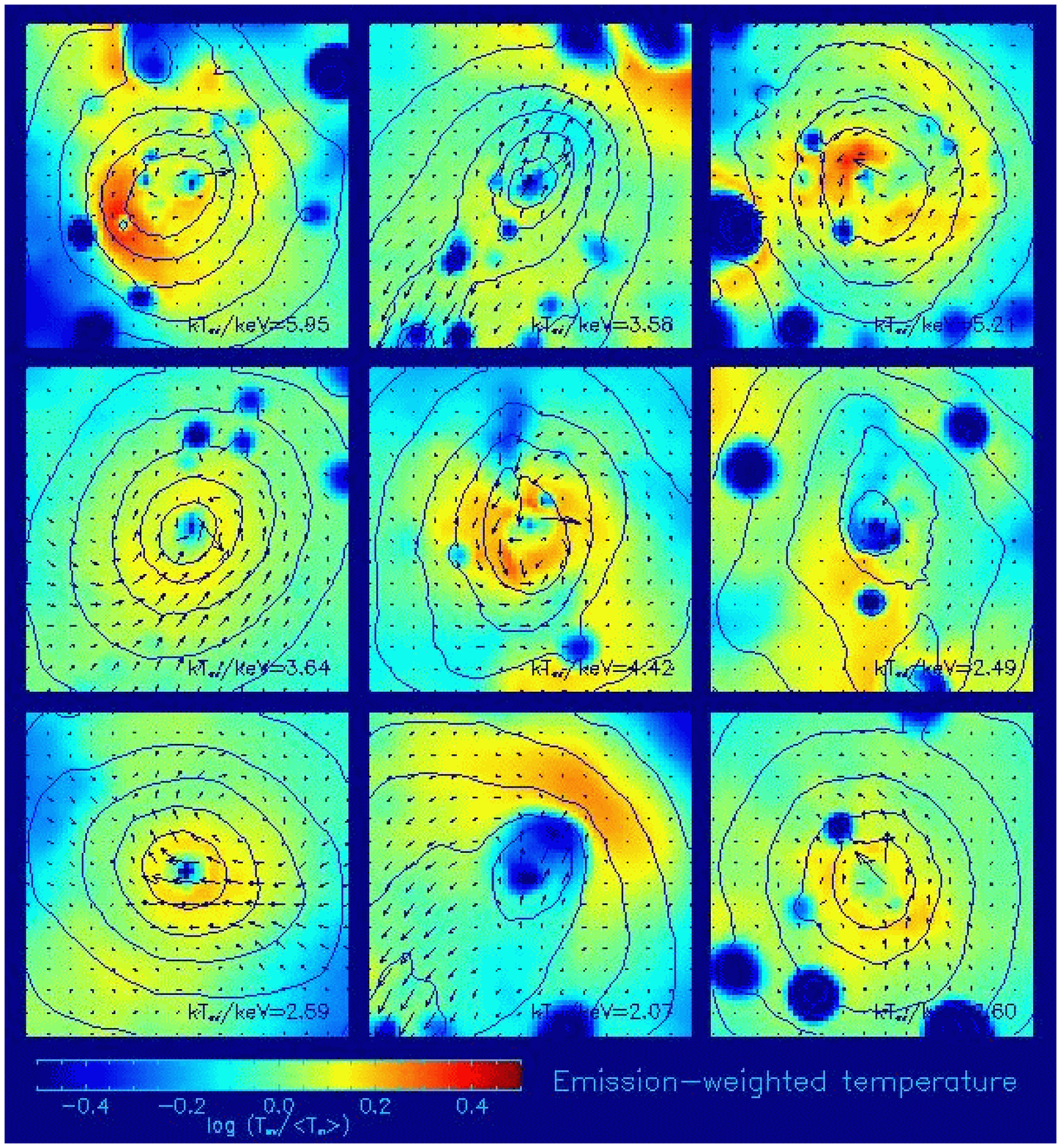,width=17.0cm,angle=0}}
\caption{Temperature maps for the {\it Radiative} run for the same
clusters as in Fig.~1}
\label{fig:radmap}
\end{figure*}

Figs.~\ref{fig:phmap} \& \ref{fig:radmap} show the emission-weighted 
temperature maps (with superimposed X-ray surface brightness contours 
and projected velocity vectors) for the {\it Preheating} and 
{\it Radiative} runs respectively. Surface brightness contours
are normalized to the maximum value, with each contour representing
variation by an order of magnitude. Velocities (in km/s) are mass-weighted 
and the length of each velocity vector represents the tangential speed (i.e.
across the map plane) of the gas at that point.

Positive temperature fluctuations are evident of up to 3 times
the average, associated with compression of the gas. There also 
exists bright clumps of cool gas, with temperatures less than half 
the mean value. These features are far more abundant in the 
{\it Radiative} maps than in the {\it Preheating} maps because the
energy injection in the {\it Preheating} run was large enough to
erase a significant amount of substructure.

Among the 9 clusters shown in Fig.~\ref{fig:phmap}, several show clear 
bimodal structure (clusters 1, 2 and 8).  This is most clear for the 
2$^{\rm nd}$ and 8$^{\rm th}$ maps, in each of which 2 cold clumps appear to
be moving away from one another.  In the {\it Radiative} maps 
(Fig.~\ref{fig:radmap})
the bimodal structure of map 1 is not so clear.  The 8$^{\rm th}$
cluster shows evidence of a sharp boundary at the leading edge of the
cold clump in the upper right quadrant; when the maps for this cluster
are re-centred to show the cold clump in the lower right of the map, a
similar sharp edge can be seen. These features are indicative of
observed cold fronts and this is investigated in the next section.

The velocity vectors superimposed on the cluster maps show clear
evidence for rotation in some cases (for example maps 3, 5 and
9). Ritchie \& Thomas (2001) found similar rotation of the velocity
field in their simulated off-centre merger of two equal mass
clusters.  Except for map 9. where the surface brightness contours are
essentially round, the contours are generally elliptical, again as
found for merging clusters. The spiral patterns seen in some of the
temperature maps (e.g. 3 \& 5) may be due to heating by shocks
produced during the merger as was seen in the simulations of Ritchie
\& Thomas during the interaction of cluster cores.  With its rounder
surface brightness contours, cluster 9 may be in a later, more relaxed
state.

\begin{figure*}
\centerline{\psfig{figure=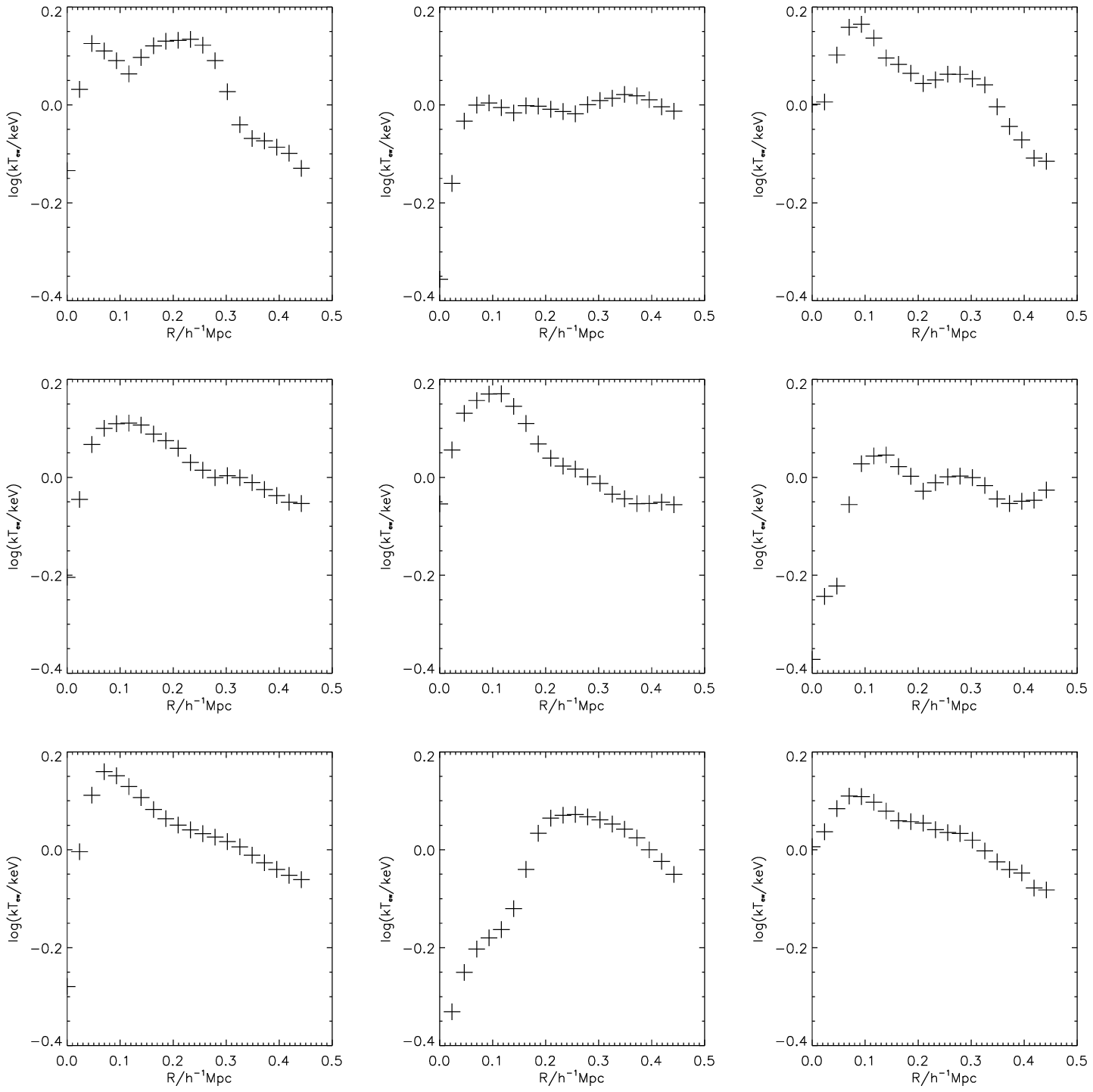,width=16.5cm,angle=0}}
\caption{Circularly-averaged temperature profiles for the {\it Radiative}
maps}
\label{fig:temprof}
\end{figure*}

The centres of all clusters show a decline in temperature. This is in
contrast to the simulated temperature maps of Loken et~al.~(2002),
produced using an Adaptive Mesh-Refinement code, where the radial
temperature profiles continue to rise at small radii.  The Santa
Barbara cluster comparison project (Frenk et al., 1999) found the
general result that for non-radiative gas, SPH codes produce a central
flat or slightly declining temperature profile, while all of the grid
codes produce a temperature profile which continues to rise to the
resolution limit.  This conflict between results using the different
codes still needs to be resolved.  However, Fabian (2002) noted that
all of the cores mapped so far by Chandra with radiative cooling times
of a few Gyr show significant central temperature drops. This
temperature drop of a factor of about 3 or more in the central region
of the clusters was explained by Fabian as possibly due to a
combination of radiative cooling and gas introduced from
dense cooling subclusters, as we see here. 

We present circularly-averaged emission-weighted temperature profiles
for the {\it Radiative} maps in Fig.~\ref{fig:temprof}. The more regular of our
clusters (4, 7, 9) exhibit the type of gradually declining temperature
profile found by Loken et al.~(2002). The others exhibit a variety of
profiles with the irregular clusters (2, 6) even appearing
approximately ishothermal when circularly averaged. It therefore seems
doubtful that a universal temperature profile can be applied to
galaxy clusters, at least in the inner regions (out to about 500
$h^{-1}$ kpc, approximately half the virial radius).

\section{Identification of a possible cold front in map 8}

\begin{figure*}
\mbox{
\psfig{figure=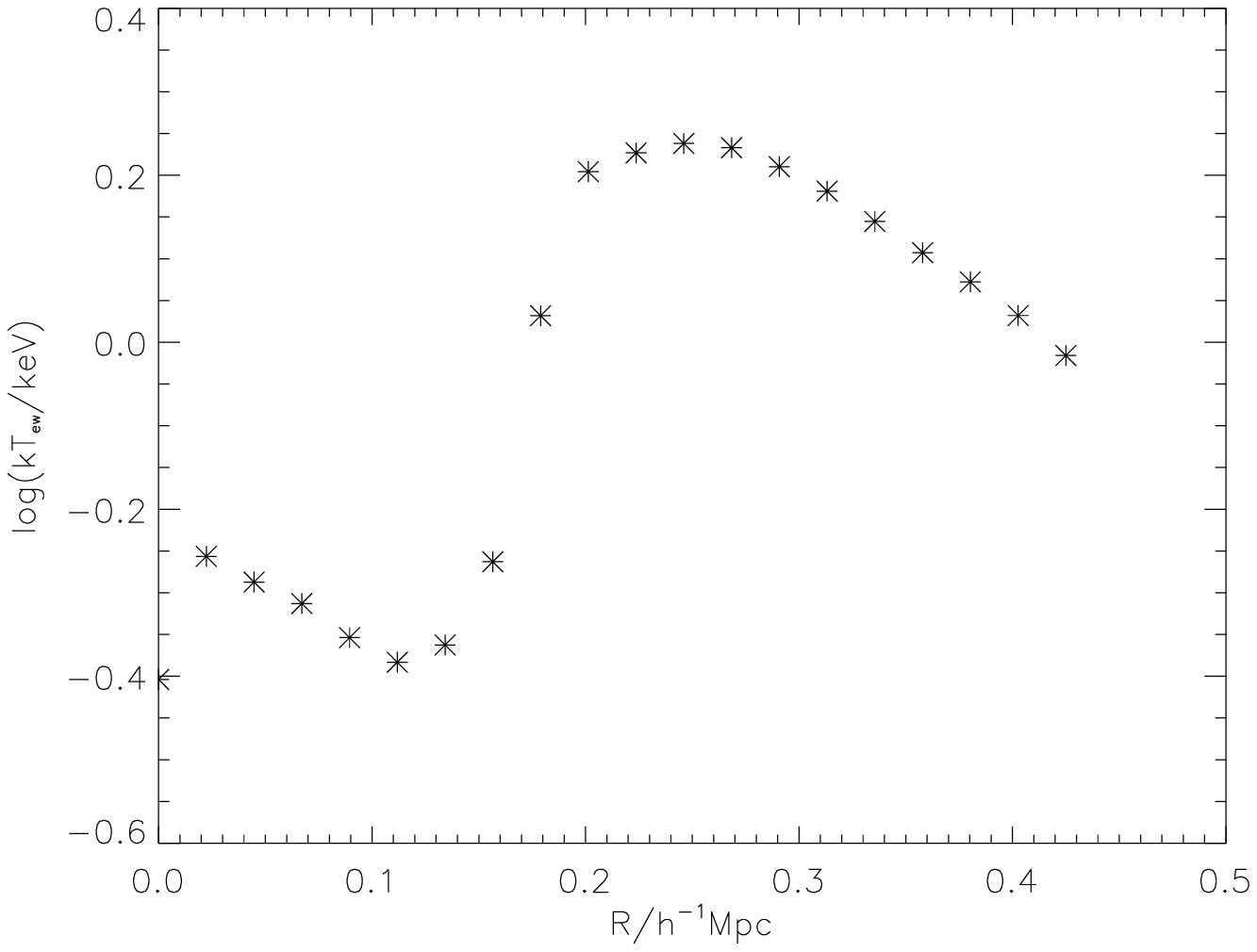,width=8.5cm,angle=0} 
\psfig{figure=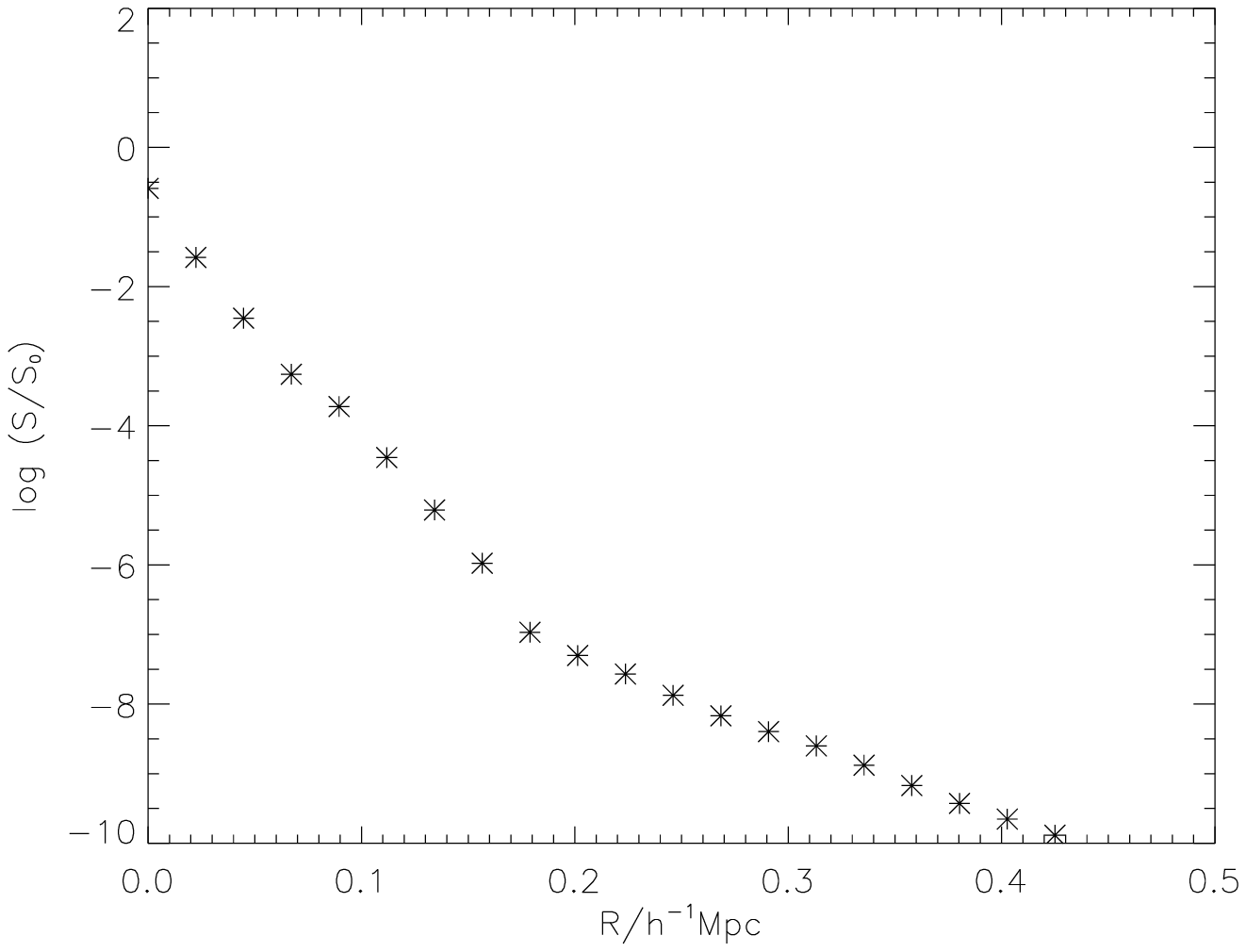,width=8.5cm,angle=0}}
\mbox{ 
\psfig{figure=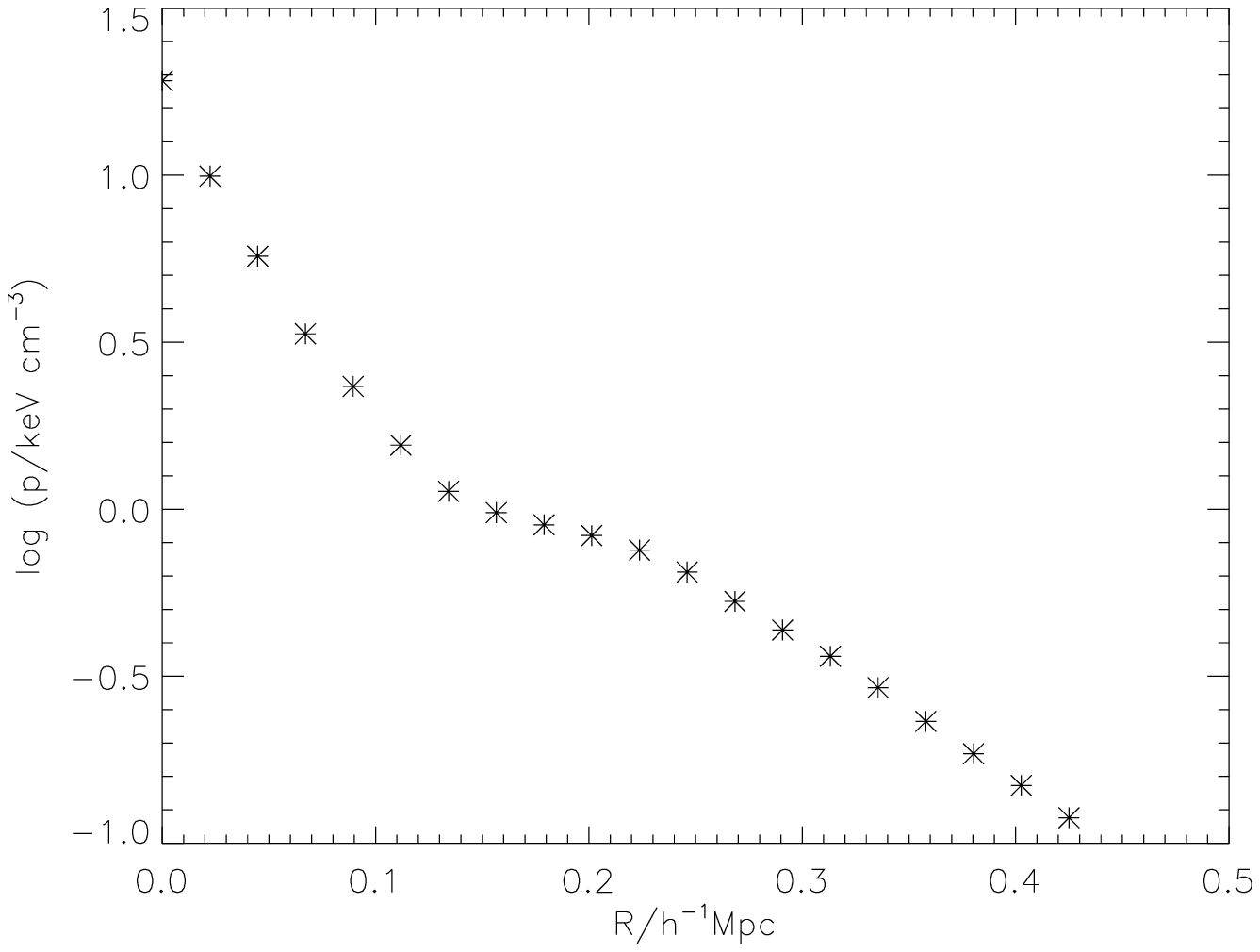,width=8.5cm,angle=0} 
\psfig{figure=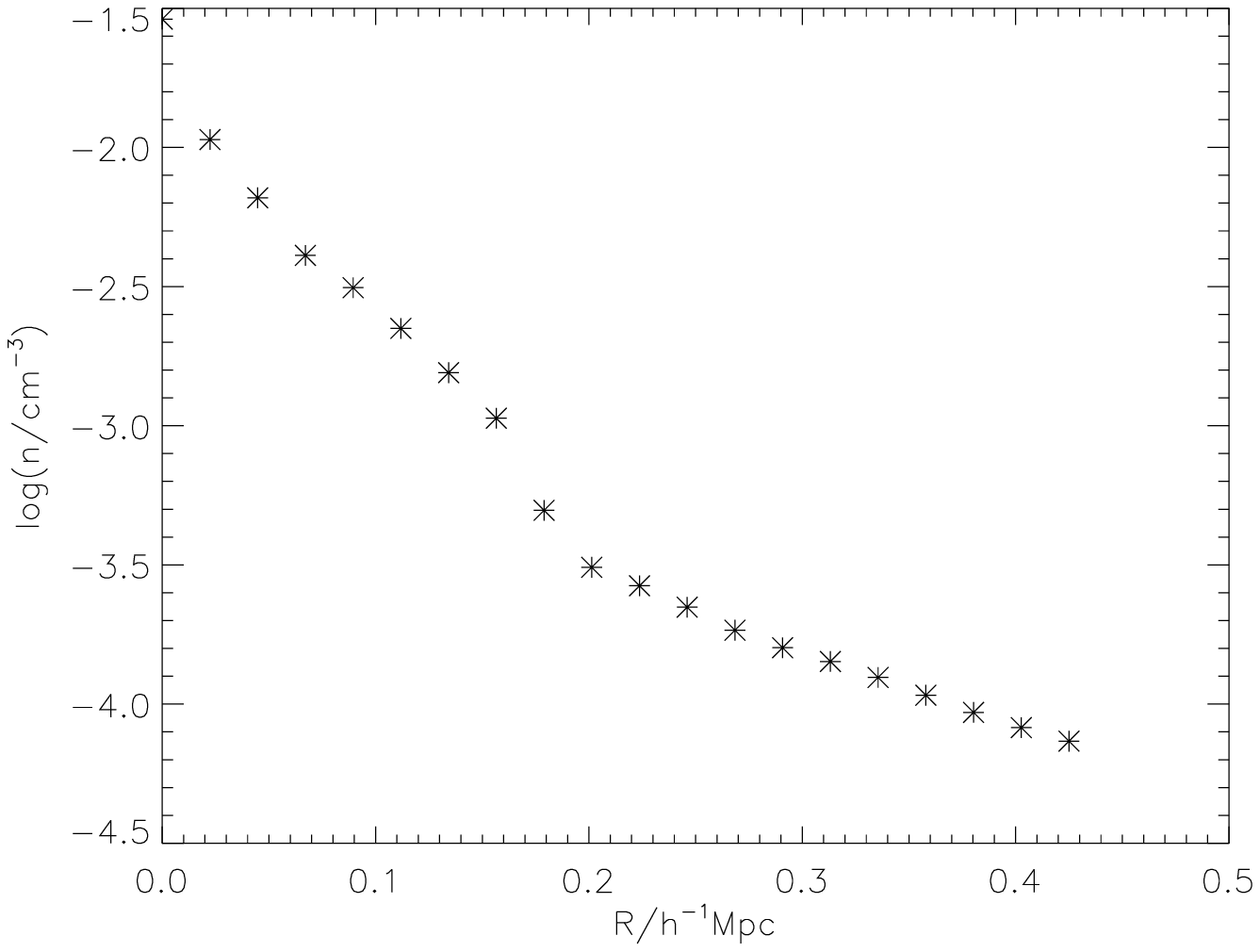,width=8.5cm,angle=0}} 
\caption{Profiles of emission-weighted temperature, $kT_{ew}$, X-ray
surface brightness, $S/S_0$, pressure, $p$, and gas density, $n$.} 
\label{fig:4prof}
\end{figure*}

Of the nine cluster maps produced here, map 8 appeared to have the
strongest visual evidence for the presence of cold fronts. To
investigate whether the physical properties are consistent with a cold
front, profiles were produced of emission-weighted temperature, X-ray
surface brightness, pressure and density across the surface brightness
and temperature discontinuity in the upper right quadarant of the {\it
Radiative} map.  The pixels with position angles between 30$^\circ$
and 60$^\circ$ with respect to the image centre were binned as a
function of radius for each of the four parameters. The profiles are
shown in Fig.~\ref{fig:4prof}.

A sharp rise in temperature by a factor of about 3 over a radial
distance of $ < 0.2 h^{-1}$ Mpc was found, corresponding to a sudden
change in the slope of the surface brightness profile.  Comparison
with the observed temperature profile across the northwestern edge of
the X-ray image for Abell 2142 in Markevitch et al.~(2000) shows that
the radial scale is very similar and the temperature for both
increases by approximately the same factor. In the region of the cold
front, the pressure profile of the simulated cluster was smooth and
fairly flat, falling off at larger radius. For the pressure profile to
flatten in this region, where the temperature is rising sharply,
implies that the gas density must be dropping sharply, again similar
to the profile in Markevitch et al. (2000).  The radial velocity of the
gas in the cold clump is highly subsonic with respect to the gas
to the right of the cold front.

Similar profiles, consistent with cold fronts, were found when the
binning was repeated across the surface brightness edge in the lower
left quadrant, and also for the {\it Preheating} run for the same
cluster. The overall picture for this cluster, taking into account
also the direction of the velocity vectors, is of two cold subclumps
moving away from each other which fits in with the merger scenario
suggested by Markevitch et al.~(2000) of subclumps falling in along
filaments. The cold cores are undisrupted by the merger and continue
on through the cluster.

\section{Discussion} 
Temperature maps of the 9 largest clusters in the simulated cluster
catalogues of M2002 show clearly that the clusters do contain
substructure.  Features that are seen include cold clumps, some of
which may be consistent with the infall of cold subclusters, hot
spiral features consistent with shocks produced in mergers, and at
least one map containing cold fronts.  Elliptical surface brightness
contours are the norm, and rotation as expected in off-centre mergers
is common. The {\it Preheating} maps appear 'cleaner', with fewer
cold clumps present than for the {\it Radiative} maps, since the gas
has been heated at z=4, wiping out many of the colder blobs. The
remaining cold clumps seen in these maps did seem to be associated
with motion into or through the cluster.

In this preliminary paper our main aim has been to demonstrate that
simulated maps contain a variety of substructure. The scale of the
structure is similar to that seen observationally and so gives some
indication of the features which should be observable. Future work
will focus on the evolution of structure in clusters and on a larger
range of cluster masses extending down to the size of groups.

\section{Acknowledgements}

The simulations used in this paper were carried out on the Cray T3E at
the EPCC as part of the Virgo Consortium programme of investigations
into the formation of structure in the Universe.  LIO was supported in
part by the Leverhulme Trust and STK by PPARC; PAT is a PPARC Lecturer
Fellow.

\end{document}